\documentclass{nature}
\usepackage[T1]{fontenc}
\usepackage[latin1]{inputenc}
\usepackage{graphicx}
\usepackage{amssymb}
\usepackage{dcolumn}
\usepackage{color}
\usepackage{bm}

\begin{document}

\title{Entangling electrons by splitting Cooper pairs: Two-particle conductance resonance and time coincidence measurements}

\author{Anindya Das, Yuval Ronen, Moty Heiblum\footnote{moty.heiblum@weizmann.ac.il}, \\Diana Mahalu, Andrey V. Kretinin and Hadas Shtrikman}
\maketitle
\paragraph{Affiliation}
Braun Center for Submicron Research, Department of Condensed Matter Physics, Weizmann Institute of Science, Rehovot 76100, Israel

Entanglement, being at the heart of the Einstein-Podolsky-Rosen (EPR) paradox, is a necessary ingredient in processing quantum information. Cooper pairs in superconductors - being composites of two fully entangled electrons - can be split adiabatically, thus forming entangled electrons. We fabricated such electron splitter by contacting an aluminum superconductor strip at the center of a suspended InAs nanowire; terminated at both ends with two normal metallic drains. Intercepting each half of the nanowire by gate - induced Coulomb blockaded quantum dot strongly impeded the flow of Cooper pairs due to large charging energy, while still permitting passage of single electrons. Here, we provide conclusive evidence of extremely high efficiency Cooper pairs splitting via observing positive average (conductance) and time (shot noise) correlations of the split electrons in the two opposite drains of the nanowire. Moreover, The actual charge of the injected quasiparticles was verified by shot noise measurements.

Two particles are said to be entangled if a measurement or a manipulation of the quantum state of one particle instantaneously affects the quantum state of the other. Hence, being non-local, the entanglement of two, separated, particles must involve simultaneous, non-local, measurements. Such two-particle state can be achieved, in principle, via particles interaction or by breaking apart a composite quantum object. For example, fully entangled photons are readily provided by low efficiency parametric down conversion of higher energy photons \cite{RosenPR1935,RogerPRL1982,MandelPRL1988}. Such a feat is not readily available for electrons. However, the closest electrical analogue to the high energy photons are Cooper pairs in a superconductor, being a natural source of entangled electron pairs. Splitting them adiabatically may give birth to entangled electron pairs. Indeed, it had been predicted and measured that Cooper pairs, emanating from a superconductor, can split into two normal metallic leads in the so called 'cross Andreev reflection' process \cite{LohneysenPRL2004,MorpurgoPRL2005,ChandrasekharPRL2006,SchonenbergerEPL2009,SchonenbergerNature2009,StrunkPRL2010,ChandrasekharNaturePhys2010,SchonenbergerPRL2011}. Such process can be conclusively verified by observing positive coincidence of arrival, or positive cross-correlation of current fluctuations, in two separated normal metallic leads that collect the split pairs \cite{MartinEPJB1999,BlatterEPJB2001,LossPRB2001,MartinPRB2011,HekkingEPL2004,MartinPRB2008}. The main difficulty in identifying such process is the overwhelming flux of Cooper pairs that enters the normal leads via 'direct Andreev reflection' (the proximity effect).
\begin{figure*}
\begin{center}
\centerline{\includegraphics[width=0.9\textwidth]{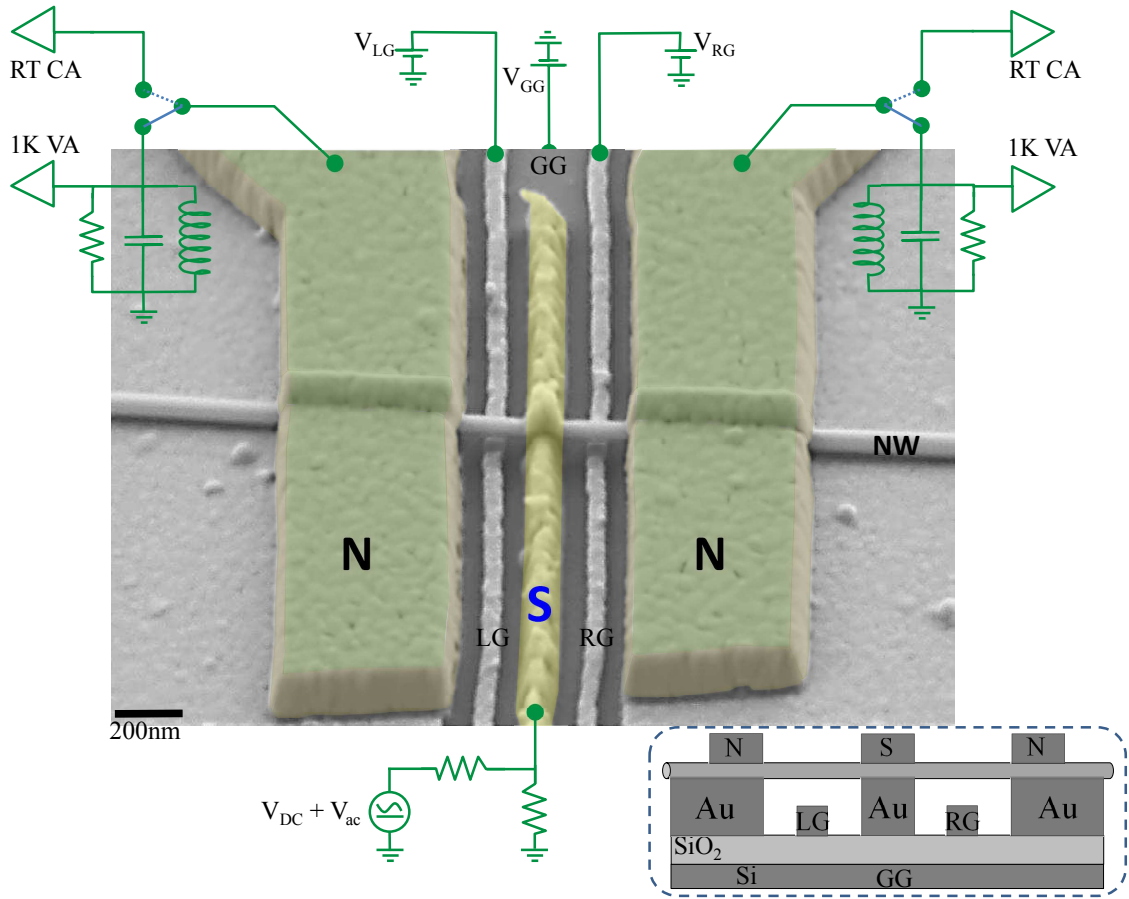}}
\caption{\textbf{Device and measurement setup.} An SEM image, with false color enhancement, of the working suspended InAs based splitting device. The nanowire is connected in its center with a superconducting $Al$ contact (S) and two normal $Au$ contacts (N) each on either side of the nanowire. Inset shows a cross-sectional schematic view. The superconducting contact is biased by a voltage source and the currents at the two normal drains are measured by room temperature current amplifiers (RT CA). Current fluctuations are measured by cold voltage amplifiers (1K VA) with an LC resonant circuits at their input. The switching between RT CA and 1K VA is done by a low temperature relay.}
\label{Figure1}
\end{center}
\end{figure*}
Such an experiment was attempted by Wei \textit{et al}, \cite{ChandrasekharNaturePhys2010}, where cross-correlation measurements were performed in an all metallic system (Al superconductor and Cu normal metal) without QDs at very low frequencies (2-6Hz) at a temperature 0.3-0.4K. The large $1/f$ noise, the relatively high temperature, and a dominant Cooper pair transport compromised the obtained data. Replacing each of the normal metallic leads with a quantum dot (QD) in the Coulomb blockade regime, done by Hofstetter et. al. \cite{SchonenbergerNature2009} indeed suppressed Cooper pairs transport \cite{LossPRB2001}, but lacked to prove coincidence splitting. Here, we provide results of coincidence measurements via observing positive cross-correlation of current fluctuations; being also reinforced by simultaneous non-local conductance measurements on both sides of the nanowire. Quenching superconductivity with a weak magnetic field suppressed the positive correlations. We obtained a splitting efficiency, defined as the ratio between single-electron to two-electron transport, as high as $\sim$ 100$\%$.

Figure 1 shows an SEM image of our device as well as a schematic illustration of the measurement setup.  A 50 nm diameter InAs nanowire, grown by a high purity Au-assisted MBE process \cite{HadasNanoLett2010}, was suspended on Au pillars above a conducting Si substrate coated with 150 nm SiO$_{2}$. A superconducting aluminum strip (S), $\sim$ 100 nm wide, was intimately contacted at the center of the nanowire, separating it into two equal sections, each $\sim$ 200 nm long, with two terminating gold ohmic contacts serving as drains (D). Aside from the conducting Si substrate, which served as a global gate (GG), two narrow metallic gates, some 50 nm wide, were used to tune the local chemical potential in each side of the nanowire. While the local gates (positively biased) accumulated electron puddles, the global gate (negatively biased) induced barriers on the sides of each puddle, thus forming two QDs on both sides of the superconducting contact. While currents were amplified with a room temperature current amplifier at $\sim$ 575 Hz, current fluctuations (broad band auto-correlation or shot noise) and their cross-correlation, were first filtered by an LC resonant circuit tuned to 725 kHz (bandwidth $\sim$ 100 kHz); amplified by a home-made cold (1K) preamplifier cascaded by a room temperature amplifier, and finally measured by a spectrum analyzer or an analog cross-correlation setup.

With all three gates unbiased, the InAs nanowire conducts $n$-type with an approximate electron density of 5$\times$10$^{6}$ cm$^{-1}$. The differential conductance of one side of the wire, say the left side, when the right side is pinched-off by its local gate (RG), is measured at 10 mK as a function of its gate voltage ($V_{LG}$), while the global gate (GG) is grounded. The conductance varies around $2e^{2}/h$ (Fig. 2a) - characteristic of 'Fabry-Perot type' oscillations \cite{HadasNanoLett2010}. Note, that conductance exceeding $2e^{2}/h$ for the first subband indicates presence of Andreev reflections with a barrier near the S-InAs interface (with maximum of $4e^{2}/h$). Under similar conditions, the non-linear differential conductance as a function of bias $V_{SD}$, is shown in Fig. 2b for two values of $V_{LG}$ (C and A in Fig. 2a). The gate voltage $V_{LG}$ mainly controls the barrier near the S-InAs interface, shifting the linear conductance from 'high' at point C to 'low' at point A, with a strikingly different non-linear conductance in the two points. 
\begin{figure*}
\begin{center}
\centerline{\includegraphics[width=1\textwidth]{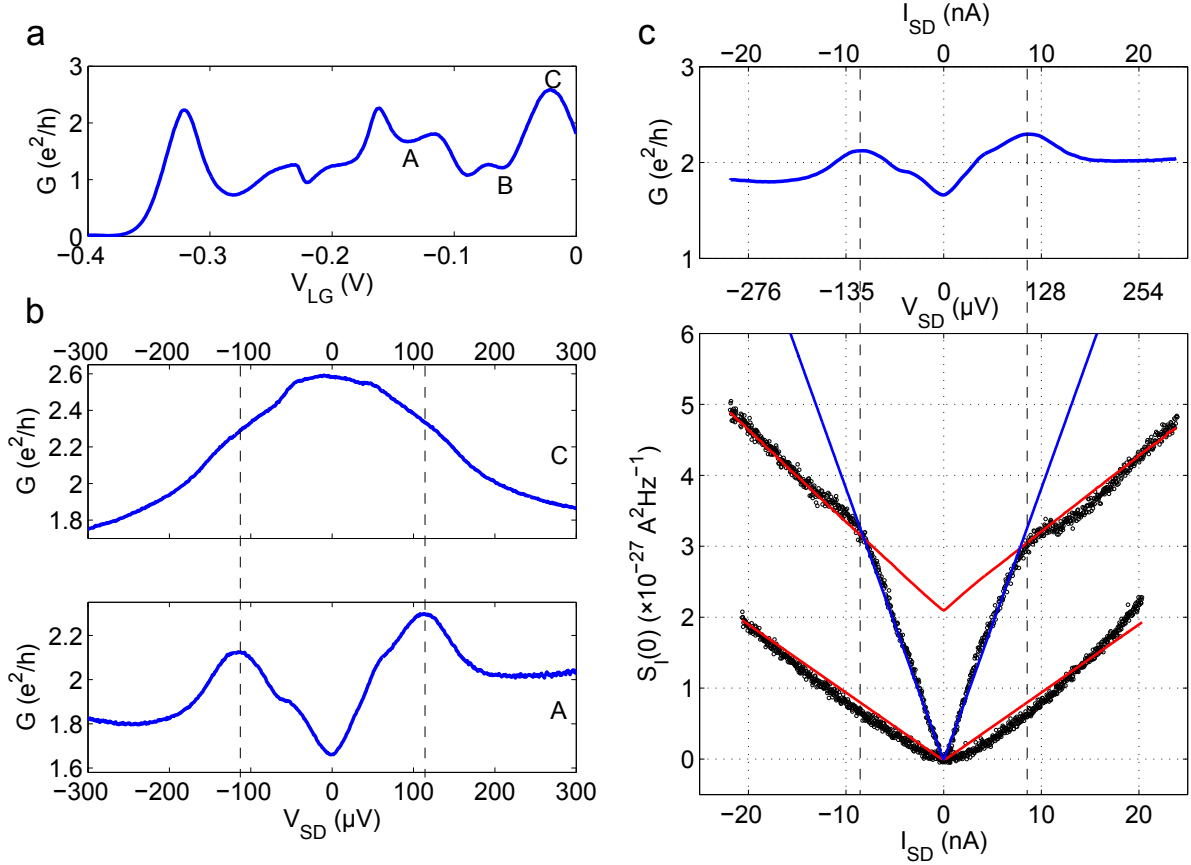}}
\caption{\textbf{Andreev reflection and charge measurement.} (\textbf{a}) $G (V_{G}))$ of the left side of the nanowire when the right side is pinched off. (\textbf{b}) Bias dependent differential conductance at points A and C. In C, the conductance is characteristic of Andreev reflection in a S-N junction, In A it is characteristic of a tunneling in a S-I-N junction (with I-barrier). Dashed lines border the  superconducting gap (2$\Delta$). (\textbf{c}) Non-linear conductance (top) and auto correlation signal (shot noise, bottom) as a function of current for $B$ = 0 and $B$ = 0.2 T. Solid lines are theoretical predictions at $T$ = 10 mK. Charge is $2e$ for $V_{SD}$ < $\Delta$ (blue line) and $e$ for $V_{SD}$ > $\Delta$ (red line).} 
\label{Figure2}
\end{center}
\end{figure*}
At point C, with linear transmission probability $t$$^{*}$ = 2.6/4 = 0.65, the conductance drops with bias - as expected for a diminishing tunneling probability of Cooper pairs as the bias approaches half the superconducting gap ($\Delta$). Alternatively, at point A with $t$$^{*}$ = 0.4, the conductance increases with bias and peaks at $V_{SD}$ = $\Delta$; when single electron tunneling dominates. The superconducting gap 2$\Delta$ $\sim$ 220$\mu$$eV$ is noted by the dotted line. A perpendicular magnetic field quenches the non-linear differential conductance with a critical field $B$ $\sim$ 0.12T.

For the above conditions, with the barrier at the S-InAs interface, the injected current is carrying shot noise, which depends linearly on the injected current ($I$) and the tunneling charge ($e^{*}$) \cite{MotyNature1997,MaillyNature2000,QuirionPRL2003,TakayanagiPRB2005}. The 'low frequency' spectral density of the 'excess noise' (shot noise above the Johnson-Nyquist and environment noise) in the single InAs channel takes the form: $S_{i}(0) = 2e^{*}I(1-t^{*})F(T)$, with $t^{*}=t$ for electrons and $t^{*}=t^{2}$ for Cooper pairs, and $F(T)=coth\zeta - 1/\zeta$, with $\zeta = e^{*}V_{sd}/k_{B}T$. Determination of the non-linear conductance (Fig. 2b-bottom, 2c-top) is crucial for accurate excess noise value since it affects the background noise (composed of thermal and 'current noise' of the preamplifier \cite{MotyNature2010}). In Fig. 2c (bottom) we plot $S_{i}(0)$ as a function of $I$ for zero magnetic field $B$ and for $B$ $\sim$ 0.2T. The blue and red solid lines are the 10 mK predictions for $t^{*}=0.4$, $e^{*}=2e$ and $t=0.63$, $e^{*}=e$, respectively; demonstrating an excellent quantitative agreement with the data (blue circles). The distinct change of slope (from $e^{*}=2e$ to $e^{*}=e$) nicely corresponds to $\Delta$ that was deduced from the conductance (Fig. 2b-bottom). A perpendicular small magnetic field ($\sim$ 0.2T) quenched the superconductivity with the excess noise nicely agreeing with $e^{*}=e$ across the full biasing range.

\begin{figure*}
\begin{center}
\centerline{\includegraphics[width=\textwidth]{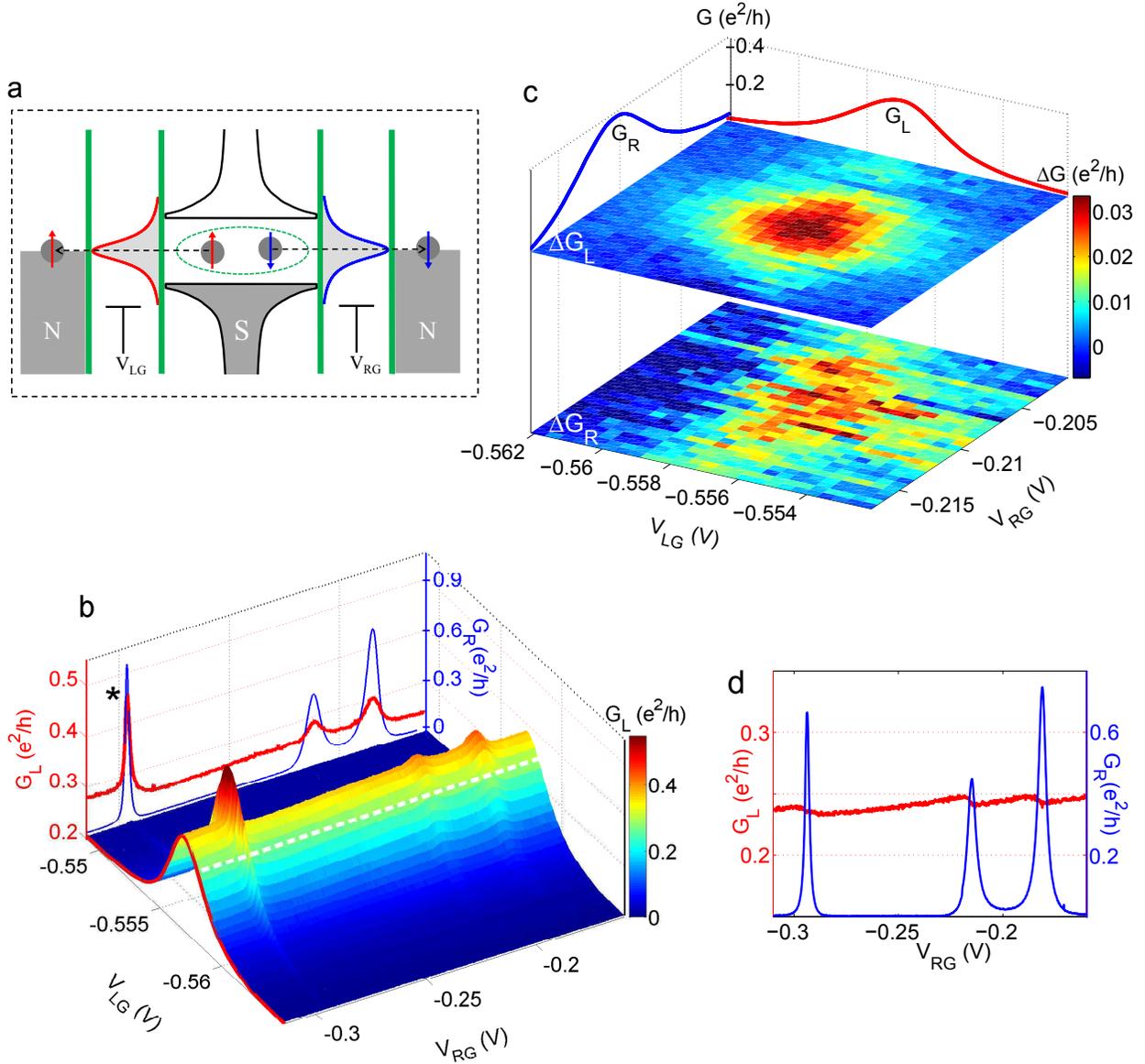}}
\caption{\textbf{Non-local conductance measurement.} (\textbf{a}) Band diagram of the system aligned for maximum Cooper pair splitting. In the non-local measurement we look at the conductance of one side of the nanowire (with a embedded QD) as a function of local gate voltage applied to the QD on the other side. (\textbf{b}) Color plot of the conductance $G_{L}$ by scanning the $V_{RG}$ for different fixed values of $V_{LG}$. The solid red line towards the left side of the color plot is a Coulomb blockade peak of QD$_{L}$ due to Cooper pairs transport when the right side is blocked. The projected red line on the top screen is the non-local conductance $G_{L}$ measured as a function of $V_{RG}$ (for the white dashed line at $V_{LG}$= -0.558V), with peaks enhanced in a corresponding manner to the conductance peaks of QD$_{R}$ (blue line on the top screen). (\textbf{c}) 2D color plot of currents due to Cooper pair splitting through the left QD ($\Delta$$G_{L}$) and the right QD ($\Delta$$G_{R}$ - scaled up by $\times$2). The red and blue lines at top left and top right wall are the Coulomb blockade peaks of QD$_{L}$ and QD$_{R}$ near $V_{LG}$ = -0.557 V and $V_{RG}$ = -0.21 V, respectively. (\textbf{d}) Non-local signal of $G_{L}$ (red line) as a function of $V_{RG}$ for $V_{LG}$ = -0.558 V at $B$ = 0.2 T. The blue line is the local $G_{R}$ as a function of $V_{RG}$.} 
\label{Figure3}
\end{center}
\end{figure*}

We now turn to study the efficiency of Cooper pairs splitting. Introducing a Coulomb blockaded QD in each side of the nanowire is expected, under suitable conditions, to suppress Cooper pair transport due to the dot's relatively large charging energy $U$. Preventing single electron injection from the superconductor necessitates, $eV_{SD}, k_{B}T < $ $\Delta$, while for quenching Cooper pair transport through the QD $eV_{SD} < U$. The characteristic energies of each QD is determined by measuring the non-linear differential conductance as a function of the DC bias ($V_{SD}$) and the local gate voltage via the so-called 'diamond' structure. We estimated the average charging energy at $U = 8-10 meV$ and the single particle level broadening at $\Gamma$ $\sim$ 200-300$\mu$$eV$. Under these conditions, with $U > $ $\Delta$ but $\Gamma \approx \Delta$, two-sequential-electron transport, proportional to $(\Gamma/\Delta)^{2}$, is barely suppressed \cite{LossPRB2001}. The efficiency, defined as the ratio of splitted current/Cooper pair current, $\eta = I_{e}/I_{CP}$, can be expressed as $\eta =$ $\frac{2\epsilon^{2}}{\Gamma^{2}}$$\times$$\frac{sin^{2}(k_{F}\delta{r})}{(k_{F}\delta{r})^{2}}$$e^{-\frac{2\delta{r}}{\pi\xi}}$, with$\frac{1}{\epsilon} = \frac{1}{\pi\Delta} + \frac{1}{U}$, $\xi$ the coherence length of a Cooper pair, $k_{F}$ the Fermi wave vector, and $\delta$$r = r_{1}-r_{2}$ the distance between the emerging split electrons; with all values related to the proximity region in the InAs \cite{LossPRB2001}.  Since $\delta$$r$ can be any value smaller than the superconductor width, we believe that its suppression factor is not important, leading, in our system to $\eta$$ = (\pi\Delta/\Gamma)^{2} \sim 1$.

We begin with non-local conductance measurements by forming two QDs in both sides of the nanowire. Applying a large negative voltage to the global gate ($V_{GG} = -15V$), while keeping the local gates ($V_{LG}$ $\&$ $V_{RG}$) at small negative voltage, induces two barriers surrounding each of the two electron puddles.  Starting, with the left QD$_{L}$, and the right side of the nanowire blocked, the conductance peaks as function of $V_{LG}$ are solely due to Cooper pairs transport. When the right side of the nanowire is also allowed to conduct split Cooper pairs transport can take place, thus enhancing also current in the left side. The highest one-electron transport at both sides is expected when the two QDs are at resonance (Fig. 3a) \cite{LossPRB2001}. Such non-local conductance measurement is shown in Fig. 3b. We simultaneously measure the conductance of both sides of the nanowire by two individual current amplifiers. In Fig. 3b, a color plot of $G_{L}$ is plotted by scanning the $V_{RG}$ for different fixed values of $V_{LG}$. The solid red line towards the left side of the color plot as a function of $V_{LG}$ is the measured $G_{L}$ of one Coulomb blockade peak due to Cooper pairs transport through QD$_{L}$ when the right side is blocked. The blue line on the top screen is the local conductance of QD$_{R}$ as a function of $V_{RG}$ when the left side is blocked. Tunning to $V_{LG}$ = -0.558V (the dashed white line in Fig. 3b) and scanning $V_{RG}$ leads to the non-local $G_{L}$ (projected red line). Conductance is enhanced by as much as 0.18 $e^{2}/h$ (marked by a star in Fig. 3b), corresponding to the conductance peaks of QD$_{R}$ with $V_{RG}$. The efficiency of splitting, defined as $\Delta$$G/G$, is proportional to $t_{R}/t_{L}$ for the left side. For tuning $V_{LG}$ = -0.558V, the efficiency is $t_{R}/t_{L}$ $\sim$ 0.7 (0.18/0.26) and is more than 100$\%$ when $V_{LG}$ is set to off resonance. Similarly, a non-local enhancement of $G_{R}$ induced by $V_{LG}$ takes place. A full representation of the non-local conductance, Cooper pair splitting currents, between the left QD (Coulomb blockade peak at $V_{LG}$ = -0.557V) and the right QD (Coulomb blockade peak at $V_{RG}$ = -0.21V) is shown in the two color plots in Fig. 3c; with $\Delta$$G_{L} (V_{LG}, V_{RG})$ (top representation) and  $\Delta$$G_{R} (V_{LG}, V_{RG})$ (bottom representation). Note that the apparent $\Delta$$G_{L}$ is bigger than $\Delta$$G_{R}$. In the presence of magnetic field ($\sim$ 0.2T), the superconductivity quenches and the non-local conductance diminishes (Fig. 3d). The residual non-local conductance, in a form of a weak saw-tooth like dependence, is the familiar 'detection behavior' of electron occupation in QDs \cite{HascoPRL1993}. Due to the proximity between the two dots ($\sim$ 300nm), the left dot senses the potential swing in the right dot when an electron is added to it, thus affecting its conductance.

\begin{figure*}
\begin{center}
\centerline{\includegraphics[width=\textwidth]{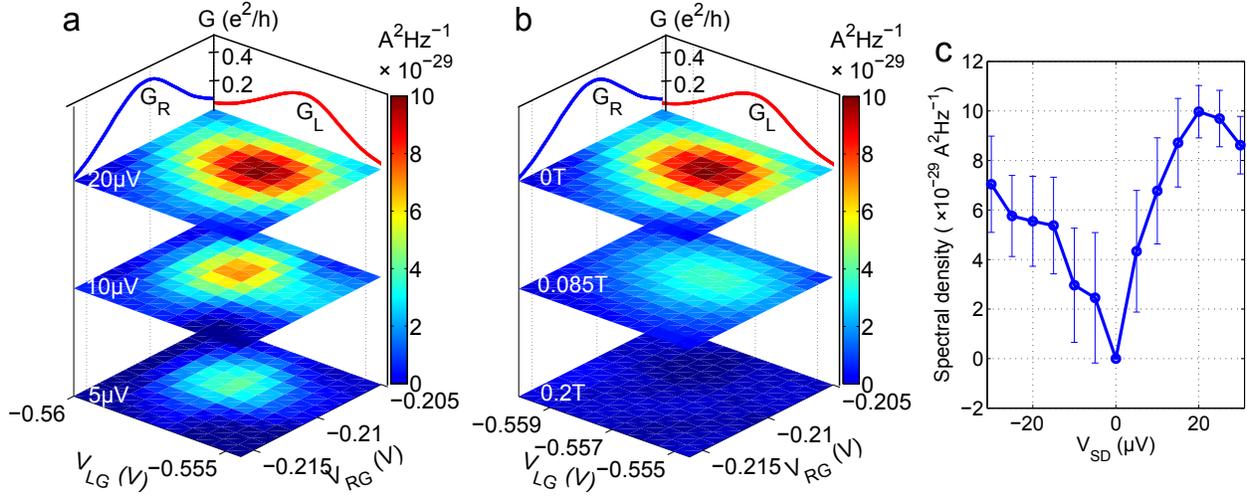}}
\caption{\textbf{Noise cross correlation.} Cross-correlated noise measured by a cross correlation setup. The current fluctuations from the normal contacts are filtered by two individual LC circuits with a matched resonant frequencies at 725 kHz with the amplified signals (by cold amplifiers) fed into an analog signal multiplier (at 725 kHz with band width of 100 kHz). (\textbf{a}) Positive cross-correlation signals for $V_{SD} =$ $20, 10$ and $5$ $\mu$$V$ between left side (Coulomb blockade peak of the QD$_{L}$ at $V_{LG}$ = -0.557V) and right side (Coulomb blockade peak of the QD$_{R}$ at $V_{RG}$ = -0.21V). (\textbf{b}) Cross-correlation signal for $B$ = 0, 0.085 and 0.2 T ($V_{SD}$ is kept at $20 $$\mu$$V$). (\textbf{c}) $S_{CC}$ as a function of $V_{SD}$ measured when both dots are at resonance.} 
\label{Figure4}
\end{center}
\end{figure*}

According to our model the non-local conductance is expected to be proportional to $t_{L}$$t_{R}$, and $\Delta$$G_{R}$ = $\Delta$$G_{L}$; which was not observed. While we do not understand the reason for this discrepancy, it might be related to a reduction in the two-electron transport, be it Cooper pairs or sequential two-electron transport, accompanying the single electron transport. Near resonance of the QD, when charge fluctuations and thus wide frequency range potential fluctuations are dominant, they can partly dephase the neighboring dot (see such saw-tooth in Fig. 3d); thus possibly affecting the higher order two-electron transport.

Measuring positive cross-correlation of current fluctuations in the two drains, assuring coincidence 'clicks', provides direct test for the existence of split Cooper pairs (since the Andreev reflections on each side are uncorrelated). To measure the cross-correlation, the current fluctuations were first amplified by a home-made cooled preamp, with the amplified signals fed to an analog signal multiplier at 725 kHz. Starting with an unbiased device, the uncorrelated background noise in both drains was nulled (being only some 2-3$\%$ of the actual auto-correlated back ground noise due to cross talk. In Figs. 4a the cross-correlation, measured with the two dots around their respective resonances ($V_{LG}$ = -0.557V and $V_{RG}$ = -0.21V), is displayed for $V_{SD} = 20, 10$ and $5\mu$$V$ DC. The cross-correlation is positive and is highest when the two dots are at resonance; in full agreement with the nonlocal conductance measurement (Fig. 3c). The dependence of the cross-correlation signal on $V_{SD}$, for the two QDs at resonance, is shown in Fig. 4c. Applying a perpendicular magnetic field, $B$ = 0.2T, quenches the superconductivity and thus eliminates the (positive) correlation between the drains' current fluctuations (Figs. 4b).

The spectral density of the cross correlation signal at zero temperature is given by $S_{CC} =  <$$\Delta$$I_{L}$$\Delta$$I_{R}>$ $\cong$ $2eI_{CAR}(1-t)$, with $I_{CAR}$ the single electron current (due to cross-Andreev reflections) in one side \cite{HekkingEPL2004}. Since $I_{CAR}/I_{AR}$ $\sim$ 0.14 (0.04/0.3 seen in Fig. 3c for the resonance peaks near $V_{LG}$ = -0.557V and $V_{RG}$ = -0.21V), and $I_{AR}$ = 500$pA$ at $V_{SD}=30$$\mu$$V$, $I_{CAR}$ = 70$pA$. Hence, we expect $S_{CC} \sim 1.7\times10^{-29} A^{2}/Hz$. However, the experimentally obtained $S_{CC}$ $\sim$ $7-10\times10^{-29} A^{2}/Hz$, which is more than four times higher than the estimated value. This discrepancy may be attributed to an under-estimated value of Cooper pair splitting efficiency that is deduced from the non-local conductance measurement. 

In Conclusion, we provided direct evidence of positive correlation between spatially separated electrons emanating from a superconductor into a suspended InAs nanowire due to splitting of Cooper pairs. We provide clear data of non-local positive current correlation as well as positive cross-correlation of current fluctuations; both at two separated drains at the ends of the InAs nanowire based QD. While coherence and spin correlation were not yet measured, the large efficiency approaching $\sim$ 100$\%$ and the positive correlation already provide strong evidence of electron entanglement.


\textbf{Acknowledgements:} We thank Yunchul Chung for contributing to the cross-correlation measurement. We also thank Hyungkook Choi, Nissim ofek, Ron Sabo, Itamar Gurman and Hiroyuki Inoue for technical help. We are grateful to Christian Schönenberger and Yuval Oreg for helpful discussions. M.H. acknowledges partial support from the European Research Council under the European Community's Seventh Framework Program (FP7/2007-2013)/ERC Grant agreement $\#$ 227716, the Israeli Science Foundation (ISF), the Minerva foundation, the German Israeli Foundation (GIF), the German Israeli Project Cooperation (DIP), and the US-Israel Bi-National Science Foundation (BSF). H.S. acknowledges partial support from the Israeli Science Foundation Grant 530-08 and Israeli Ministry of Science Grant 3-66799.

\end{document}